\documentstyle[10lomcon,graphicx]{article}

\begin{document}

\title{DISCOVERY OF $\eta$-MESIC NUCLEI}

\author{G.A. Sokol\footnote{e-mail: gsokol@x4u.lpi.ruhep.ru},
        L.N. Pavlyuchenko}

\address{ P.N. Lebedev Physical Institute, Moscow, Russia }

\maketitle

\bigskip
\abstracts{
First experimental results from a photoreaction which confirm an
existence of eta-mesic nuclei are discussed. The experiment was
performed at the 1 GeV electron synchrotron of LPI.  Two values of the
end-point energy of the bremsstrahlung beam were used, 650 and 850 MeV
(below and above $\eta$-photoproduction threshold on the nucleon).
Correlated $\pi^+ n$ pairs flying from the $^{12}$C target were
detected by two TOF scintillator spectrometers placed in opposite
directions and transversely to the $\gamma$ beam.  Measured energy
spectra provide a strong evidence for a creation of eta-mesic nuclei as
intermediate states of the reaction studied.}

\section{Introduction}
A modern view of the structure of atomic nuclei as a system of protons
and neutrons appeared in 30s years of the 20th century after the
discovery of the neutron. Since then many new particles and their
resonance states were found which led in 60s to a development of the
conception of quarks.  Accordingly, a question arose of a possible
existence of nuclear bound systems including, apart from protons and
neutrons (nucleons), new particles too.

In 50s years, hypernuclei --- a new sort of nuclei consisted of
$\Lambda$ or $\Sigma$ hyperons (apart from the nucleons) --- were
discovered.  The $\Lambda$ and $\Sigma$ hyperons are close analogues of
nucleons as for their masses and the quark content. Their masses are
only bigger by $\sim 200$ MeV than the nucleon mass $(\sim 940$ MeV)
and they consist of 3 constituent quarks as the nucleons do, one of
them being however the strange ($s$) quark.

\section{$\eta N$ interaction and $\eta$-nuclei}

An idea that a bound state of the $\eta$-meson and a nucleus (the
so-called $\eta$-mesic nucleus) can exist in Nature was put forward
long ago by Peng [1].  The $\eta$-meson does not have an open
strangeness. It consists of 2 quarks (more exactly, a quark and
antiquark with the total isospin 0) of different flavors, of which
$\approx 50 \%$ is an $s \bar s$ pair. The mass of $\eta$ is 547.5 MeV,
i.e.\  about 1/2 of the nucleon mass. The suggestion of J.C.~Peng was
based on the first estimate of the $\eta N$ scattering length $a_{\eta N}$,
\begin{equation}
   a_{\eta N} = (0.27 + i \cdot 0.22)~ \rm fm,
\end{equation}
derived by Bhalerao and Liu [2] from a coupled-channel analysis of the
reactions $\pi N\to\pi N$, $\pi N\to\eta N$ and $\pi N\to\pi\pi N$.
Owing to Re$\,a_{\eta N} > 0$, there is an average attractive $s$-wave
potential between a slow $\eta$ and a nucleon. For extended nuclei,
such an attraction should be sufficient for making the $\eta$-meson
bound, provided the life-time of the $\eta$-meson in the nucleus is not
too short. A quantum-mechanical consideration done by Liu and  Haider
[3] and based on the $\eta N$ potential corresponding to Eq.~(1)
predicted that bound states of the $\eta$-meson and a nucleus $A$ exist
for all $A \ge 11$. Later on, this conclusion was strengthen.  More
sophisticated coupled-channel analysis [4,5] taking into account both
resonance and nonresonance contributions arrived to the scattering
length Re$\,a_{\eta N}$ which is about 3 times bigger than the very
first estimate (1):
\begin{equation}
  a_{\eta N} = (0.75 + i \cdot 0.29)~ \rm fm ~[5].
\end{equation}
For such a large $a_{\eta N}$, $\eta$-mesic nuclei are predicted to
exist for all nuclei with $A \ge 4$.  With slightly larger $a_{\eta N}$,
$\eta$-bound states are to be possible for $A=3$ and even for $A=2$
[6].  It is worth emphasizing that elementary $\eta N$ scattering
amplitudes are theoretically derived from other reactions like $\pi N
\to \pi N$ and $\pi N \to \eta N$ through extrapolations based on a
factorization [2] or, in the latest works, on unitarity constraints
[4,5]. Since, however, not all important channels are involved into
those extrapolations (missing are, e.g., $\pi N \to K \Lambda$ and
$\eta N \to K \Lambda)$, it is not clear how reliable are the obtained
results. A difference between Eqs.~(1) and (2) may give a hint about
real uncertainties. Therefore, experimental studies of bound states of
various $\eta A$ systems would greatly contribute to learning
elementary $\eta N$ scattering.

\begin{figure}[hbt]
\centerline{\includegraphics*[width=0.6\textwidth]{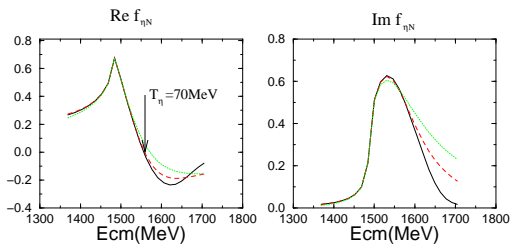}}
\vspace*{-2em}
\caption{Energy dependence of the amplitude $f_{\eta N}$
for the process $\eta N \to \eta N $ [5].}
\end{figure}

Real part of the $\eta N$ scattering amplitude $f_{\eta N}$, the
threshold value of which is just equal to the scattering length
$a_{\eta N}$, remains positive up to the kinetic energies of $\eta$
below 70 MeV [5] (Fig.~1). This means that an effective $\eta A$
attraction exists in a wide near-threshold energy region $E_{\eta}^{\rm
kin} \approx 0 \div 70$ MeV.  The attractive forces in the final state
have to lead to a near-threshold enhancement in the total and
differential cross section of real-$\eta$ production by different
beams. Such an enhancement was indeed observed in several reactions
including $p(d,{}^3{\rm He})\eta$ [7,8] and $d(d,{}^4{\rm He})\eta$
[9,10], thus supporting the existence of the $\eta A$ attraction even
for the lightest nuclei.  Nevertheless, all these experiments which
have deal with $\eta$ in the final state cannot directly prove that
bound $\eta A$ states do really exist. A well-known counter example is
provided by the $NN$ system in the ${}^1S_0$ state which has a virtual
rather than a real level and it has a negative rather than positive
scattering length.

\section{A previous experience}

Two attempts to find $\eta$-nuclei in the missing mass spectrum of the
reaction $\pi^+ A \to p X$ were performed at BNL [11] and LAMPF [12]
soon after the first theoretical suggestions [1,2]. Both the
experiments failed to find a signal of the $\eta$-nuclei, perhaps owing
to their much bigger total widths than then expected.

A new experiment started in 1994 at the 1 GeV electron synchrotron of
Lebedev Physical Institute aiming at a search for $\eta$-nucleus bound
states in photoreactions.  In 1998, the first experimental data were
obtained [13] which indicated that a bound state of the $\eta$-meson
and a nucleus $A=11$ does exist.

\section{LPI experiment}

Below we describe in some details a method used in [13] for
identification of the bound states of the $\eta$-meson and a nucleus,
as well as a procedure of measurements and data analysis.

As a preliminary remark note that the use of the photon beam for
formating $\eta$-nuclei has certain advantages when compared with the
use of the pion beam.  The reason is that

 -- photons interact with nucleons throughout the whole nuclear
volume whereas pions strongly interact only with nucleons near the
nuclear surface;

 -- high photon flux of order $10^9{-}10^{10}~\rm s^{-1}$ in the
energy range of $\Delta E_\gamma \simeq 200$ MeV provides quite an
adequate reaction rate.

\section{Method of identification of $\eta$-mesic nuclei}

An important idea of the work [13] was in searching for decay products
of $\eta$-nuclei which are $\pi N$ pairs arising from decays of the
$S_{11}(1535)$ resonance inside a nucleus [14]. A mechanism of
$\eta$-nuclei production and decay in photoreactions is schematically
shown on Fig.~2.
\begin{figure}[hbt]
\centerline{\includegraphics*[width=0.5\textwidth]{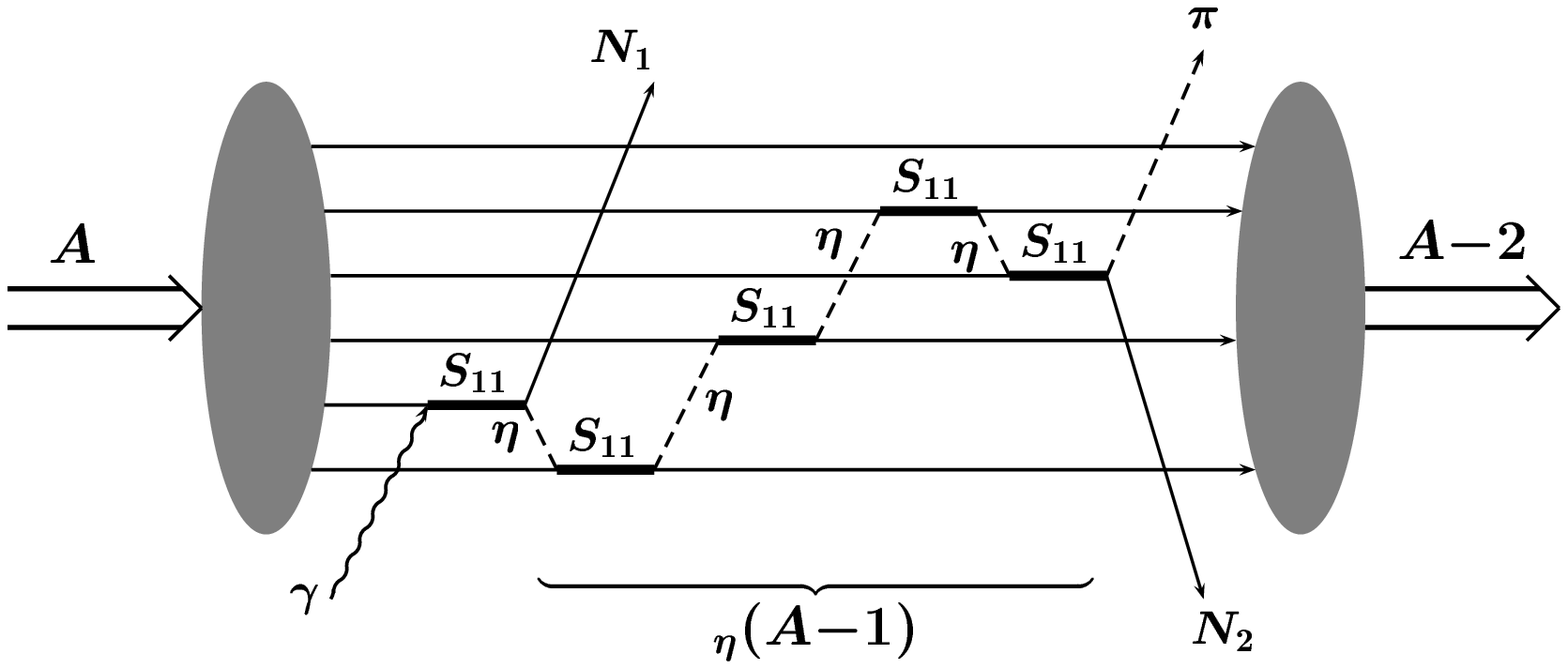}}
\caption{Mechanism of creation of a ${\eta}$-mesic
nucleus and its decay into a $\pi N$ pair in a photoreaction.}
\end {figure}
At the first stage, the incoming photon produces a slow $\eta$-meson
and a fast nucleon, the latter leaving the nucleus.  Then, in the
nuclear medium, the $\eta$-meson undergoes multiple elastic scattering,
$\eta N \to S_{11} \to \eta N \to S_{11}  \ldots$.  At the last stage,
the $\eta$-meson annihilates after interaction with a nucleon,
$\eta N\to S_{11} \to \pi N$, so that a $\pi N$ pair emerges which
leaves the nucleus too.  The $S_{11}(1535)$ resonance plays a
fundamental role in this dynamics.  It provides production and
annihilation of the $\eta$-meson and it also predetermines an
attractive $\eta$-nucleus interaction.  The chain of elastic processes
$\eta N \to S_{11} \to \eta N \to \ldots S_{11}$ in the nuclear medium
results in randomizating momenta of the $S_{11}(1535)$ resonance and
the $\eta$-meson.  Fermi motion of nucleons is important for this
randomization because the $\eta$-meson scatters on nucleons having
random velocities and directions.  The most important result of the
randomization is the isotropic momentum distribution of decaying
$S_{11}(1535)$ resonances.  It leads to an isotropic distribution of
the $\pi N$ pairs and, since the momentum of $S_{11}(1535)$ in the
nucleus is small, to the opening angle $\theta_{\pi N}$ close to
$180^\circ$.  This is a basic idea of the method of $\eta$-nuclei
identification suggested in [14]. The method consists in detection and
energy measurements of components of $\pi N$ pairs from $S_{11}$-resonance
decays in the nuclear medium. It is worth mentioning that
$\pi^+ n$ pairs flying transversely to the photon beam cannot be
produced via the one-step reaction $\gamma p \to \pi^+ n$ in the
nucleus when the photon energy is as high as $700{-}800$ MeV.

\section{Simulation of $\pi N$ events}

Theoretical estimates [13,15] show that the binding effects lead to a
full dominance of the reaction mechanism related with multiple
rescattering of $\eta$ and with a formation of the intermediate
$\eta$-nucleus over a mechanism of non-resonance (background)
production of the $\pi N$ pairs in the subthreshold region of the
invariant mass $\sqrt{s_{\pi N}} < m_\eta + m_N$ for the subprocess
$\eta + N \to \pi + N$. A peak in the mass distribution of $\pi N$ is
theoretically expected in this region.
\begin{figure}[hbt]
\includegraphics*[width=0.8\textwidth]{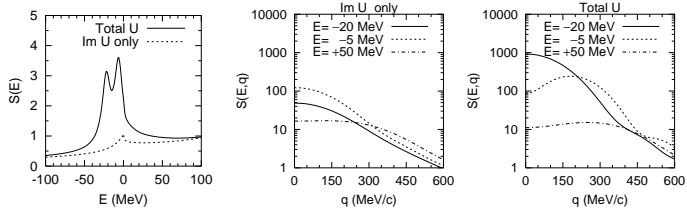}
\caption{Spectral functions $S(E)$ and $S(E,q)$ (in arbitrary units) of
the kinetic energy $E$ and momentum $q$ of $\eta$ in the nucleus.  They
are found with a rectangular-well optical potential simulating the
nucleus $^{12}$C.  For a comparison results obtained with dropping out
the attractive (i.e.\ real) part of the $\eta A$ potential $U$ are also
shown.}
\end{figure}
\begin{figure}[hbt]
\includegraphics*[width=0.35\textwidth,bb=114 560 305 769]{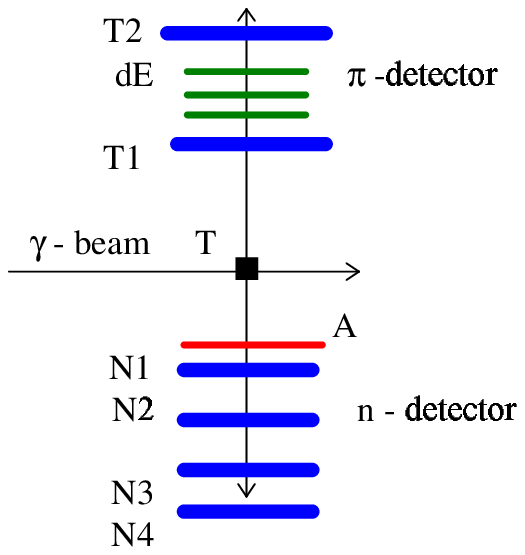}
\includegraphics*[width=0.30\textwidth,bb=000 000 184 220]{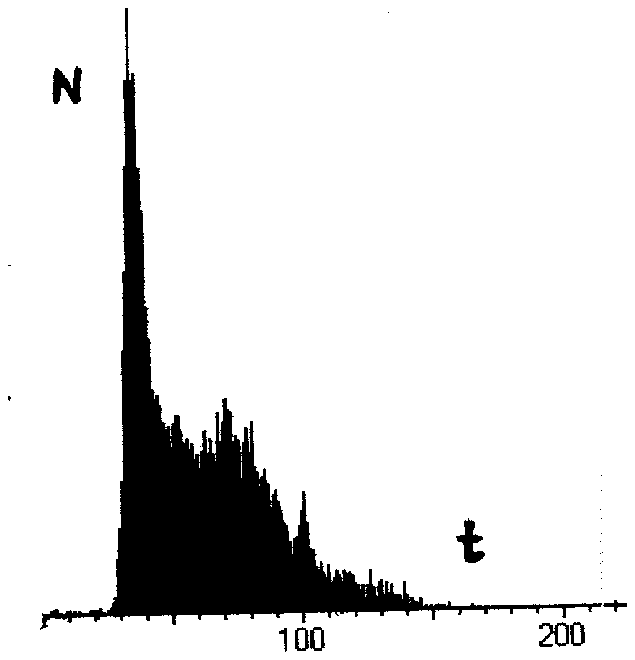}
\includegraphics*[width=0.30\textwidth,bb=000 000 200 210]{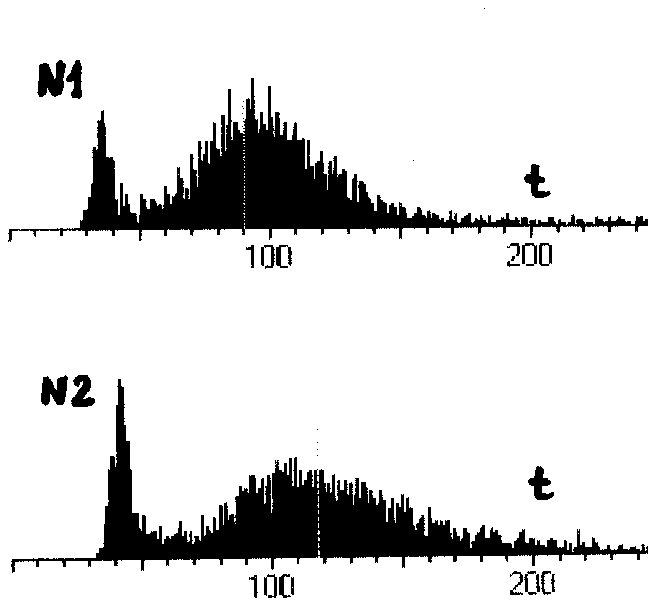}
\vspace{-2em}
\caption{Layout of the experimental setup. Shown also are the
time-of-flight spectra in the $\pi$ (left) and $n$
(right) spectrometers.}
\end{figure}
This is illustrated in Fig.~3 where the spectral function $S(E)$ of the
(kinetic) energy $E$ of $\eta$ in the nucleus is shown which is
proportional to the number of $\eta N$ collisions that the $\eta$
undergoes when travels through the nucleus. Another spectral function,
$S(E,q)$, shows a distribution of the produced $\pi N$ pairs over their
total energy $E + m_{\eta} + m_N$ and the total momentum $q$.  A
presence  of the $\eta N$ attractive potential $U$ produces strong
enhancements in these spectral functions in the energy-momentum region
corresponding to the bound $\eta A$ states.

\section{Experimental setup and procedure of measurements}

The reaction studied in our work was
\begin{equation}
     \gamma + {}^{12}{\rm C} \to p(n) +
              {}^{11}_{~\eta} {\rm B}
              ~(^{11}_{~\eta} {\rm C})
     \to \pi^+ + n + X,
\end{equation}
in which energies and momenta of of $\pi^+$ and $n$ are correlated.
Among four different $\pi N$ channels of $S_{11}(1535)$ decays
we have chosen the $\pi^+ n$ channel because:

 -- detection of the $\pi^0$-decay modes is more expensive since this
requires a use of the $4\pi$ geometry of the $\gamma$-spectrometer;

 -- detection of the $\pi^- p$ pairs with proton energies of about 100
MeV dictates a use of rather thin targets and leads therefore to a
reduced count rate.

At the same time the selected decay channel ($\pi^+ n$) allows
the use of thick targets and a rather simple setup.  Average energies
of $\pi n$ pair's components are $\langle E_\pi \rangle = 300$ MeV and
$\langle E_n \rangle = 100$ MeV.  The setup used included two arms of
time-of-flight (TOF) scintillator spectrometers (Fig.~4).  The pion
spectrometer contained two scintillator blocks of $50\times 50 \times 2$
cm$^3$ and $50\times 50\times 5$ cm$^3$. The neutron spectrometer
contained an anti-coincidence scintillator detector of the size
$50\times 50\times 2$ cm$^3$ and four scintillator detectors of the
size $50\times 50 \times 10$ cm$^3$.  Efficiency of the
anti-coincidence (veto) detector was $90 \%$.
Efficiency of the neutron detecting was about $8 \div 10 \%$. Each
scintillator detector had four photo-tubes FEU-63 placed on the edges of the
scintillator block; they were used for determination of the coordinates
of particles passed through the detector using time differences of
light signals coming to the photo-tubes.  An accuracy of the coordinate
determination was about $\Delta x = \Delta y = 1.5$ cm. An accuracy of
the determination of time was $\Delta t = 250$ ps.
\begin{figure}[thb]
\vspace{2cm}
\unitlength=1pt
\begin{picture}(100,55)(0,0)
\put(-5,90){$a)$}
\put(280,20){\vector(-3,1){60}}
\put(110,95){$(\pi^+ n)$}
\put(140,90){\line(1,-1){45}}
\put(140,90){\line(-1,-1){50}}
\put(290,10){$(\pi \pi)$}
\put(290,60){$E_{\gamma \rm max}=650$ MeV}
\put(290,40){$\theta_{\pi}=\theta_n=50^o$}
\put(285,65){\vector(-1,0){20}}
\includegraphics[width=10cm,bb=46 492 565 685]{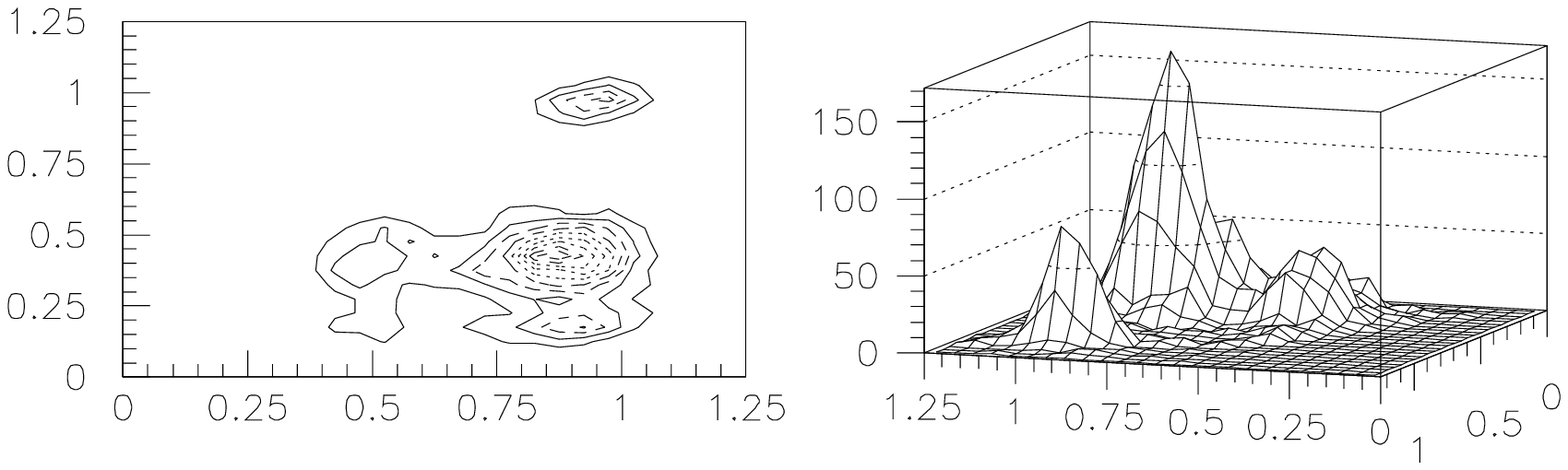}
\end{picture}
\\[1em]
\begin{picture}(100,80)(0,0)
\put(-5,90){$b)$}
\put(280,20){\vector(-3,1){60}}
\put(290,10){$(\pi \pi)$}
\put(140,90){\line(2,-3){35}}
\put(140,90){\line(-2,-1){40}}
\put(290,60){$E_{\gamma \rm max}=650$ MeV}
\put(290,40){$\theta_{\pi}=\theta_n=90^o$}
\put(130,95){$(\pi^0 \pi^0)$}
\put(285,65){\vector(-1,0){20}}
\includegraphics[width=10cm,bb=46 492 565 685]{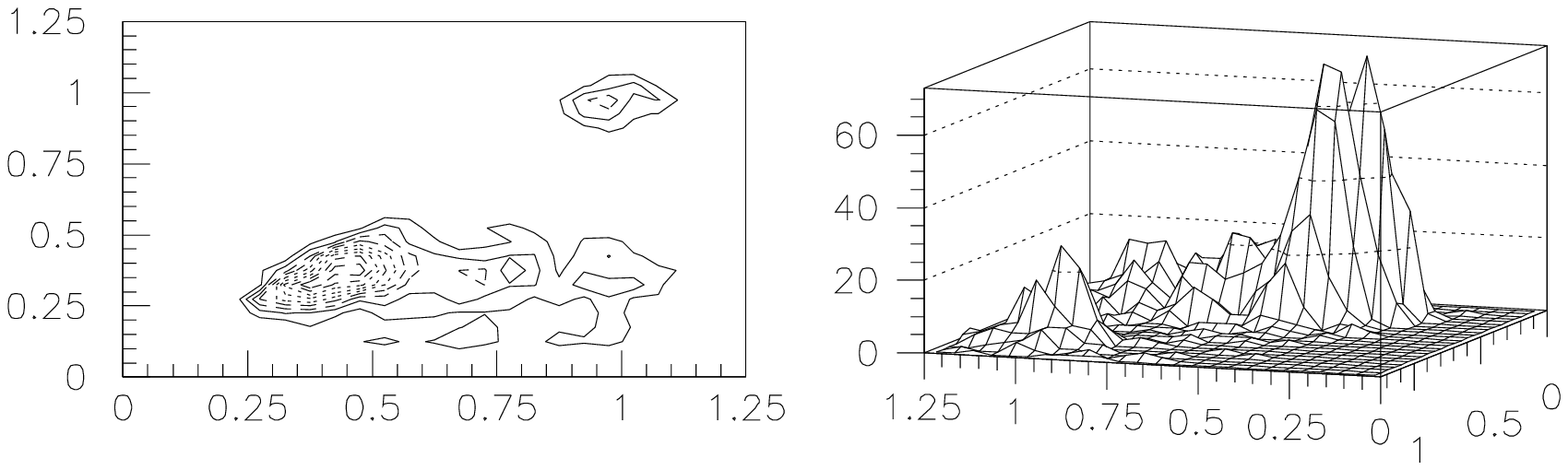}
\end{picture}
\\[1em]
\begin{picture}(100,80)(0,0)
\put(-5,90){$c)$}
\put(280,100){\vector(-3,-1){60}}
\put(290,60){$E_{\gamma \rm max}=850$ MeV}
\put(290,40){$\theta_{\pi}=\theta_n=90^o$}
\put(110,95){$(\pi ,\pi),~(\pi^+ n)$}
\put(285,65){\vector(-1,0){20}}
\put(140,90){\line(1,-1){45}}
\put(140,90){\line(-1,-1){50}}
\hspace{-0.5cm}\raisebox{1.5cm}{$\beta_n$}
\hspace{2cm}\raisebox{-0.1cm}{$\beta_\pi$}
\hspace{4cm}\raisebox{-0.1cm}{$\beta_\pi$}
\hspace{2cm}\raisebox{0.0cm}{$\beta_n$}
\hspace{-9.6cm}
\includegraphics[width=10cm,bb=46 492 565 685]{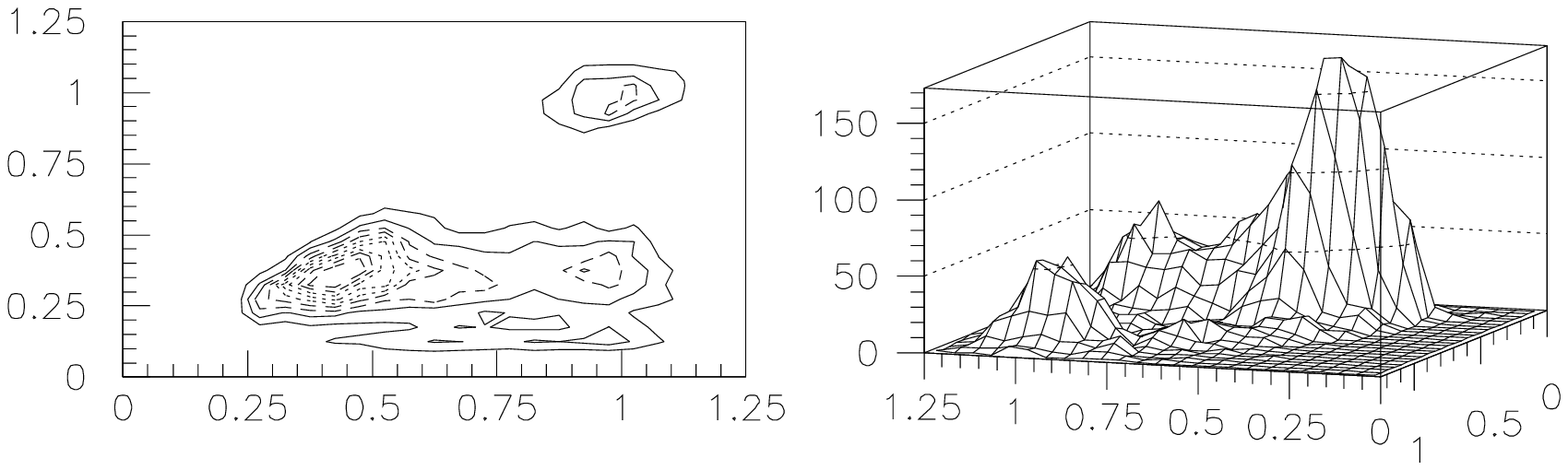}
\\[0em]
\end{picture}
\caption{Raw $\pi^+ n$-event distributions over the particle velocities
$\beta_n$ and $\beta_\pi$ for the ``calibration" $(a)$, ``background"
$(b)$ and ``effect$+$background" $(c)$ runs.}
\end{figure}
\begin{figure}[hbt]
\includegraphics*[width=0.7\textwidth,bb=72 258 528 569]{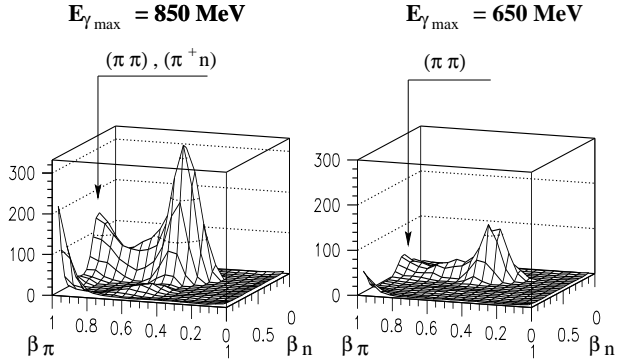}
\caption{Unfolded two-dimentional distributions over the velocities
$\beta$'s of the $\pi^+ n$ events with the end-point energy of the
bremsstrahlung spectrum $E_{\gamma \rm max} = 850$ and 650 MeV.}
\end{figure}
Measurements in the experiment were made in three different runs.  In
the ``calibration" run $(a)$ both $\pi^+$ and $n$ spectrometers were
positioned at angles
$\langle \theta_\pi \rangle = +50^\circ$ and
$\langle \theta_n \rangle = -50^\circ$
with respect to the photon beam direction and the bremsstrahlung beam
end-energy $E_{\gamma \rm max} = 650$ MeV was used.  This
configuration is suitable for measuring quasi-free single-pion
photoproduction $\gamma p \to \pi^+ n$ on protons in the nuclear
target.  It was used for final tests of both arms of the setup since
energies of the detected particles were close to expected energies of
$\pi^+ n$ pairs arising from the $S_{11}(1535)$ decays.
In the ``background" run $(b)$ the spectrometers were shifted to
$\langle \theta_\pi \rangle = +90^\circ$ and
$\langle \theta_n \rangle = -90^\circ$
and the beam energy was still $E_{\gamma \rm max} = 650$ MeV.  In such a
configuration single-pion production events were eliminated and most of
the detected events were related to a quasi-free double-pion
photoproduction $\gamma N \to \pi \pi N$.  In the last,
``effect$+$background" run $(c)$ the configuration of the setup was as
in the ``background" run $(b)$ with the except for the end-point energy
which was increased to $E_{\gamma \rm max} = 850$ MeV thus making
$\eta$ production possible.

\section{Data analysis and results}

Fig.~5 shows a two-dimensional spectrum of the detected particle's
velocities $\beta$ for all three runs.  The measured velocities
$\beta_i = L_i/ct_i$ are subject to fluctuations stemming from errors
$\delta t_i$ and $\delta L_i$ in the time-of-flight $t_i$ and the
flight base $L_i$ measured in the experiment. Such fluctuations are
clearly seen in the case when $\pi^0 \pi^0$ (actually photons or
relativistic electrons or positrons) hit scintillators. Therefore,
experimentally observed velocities are close but not equal to 1 (see in
Fig.~5).
\begin{figure}[hbt]
\centerline{
\includegraphics[width=4.0cm,bb=157 322 443 613]{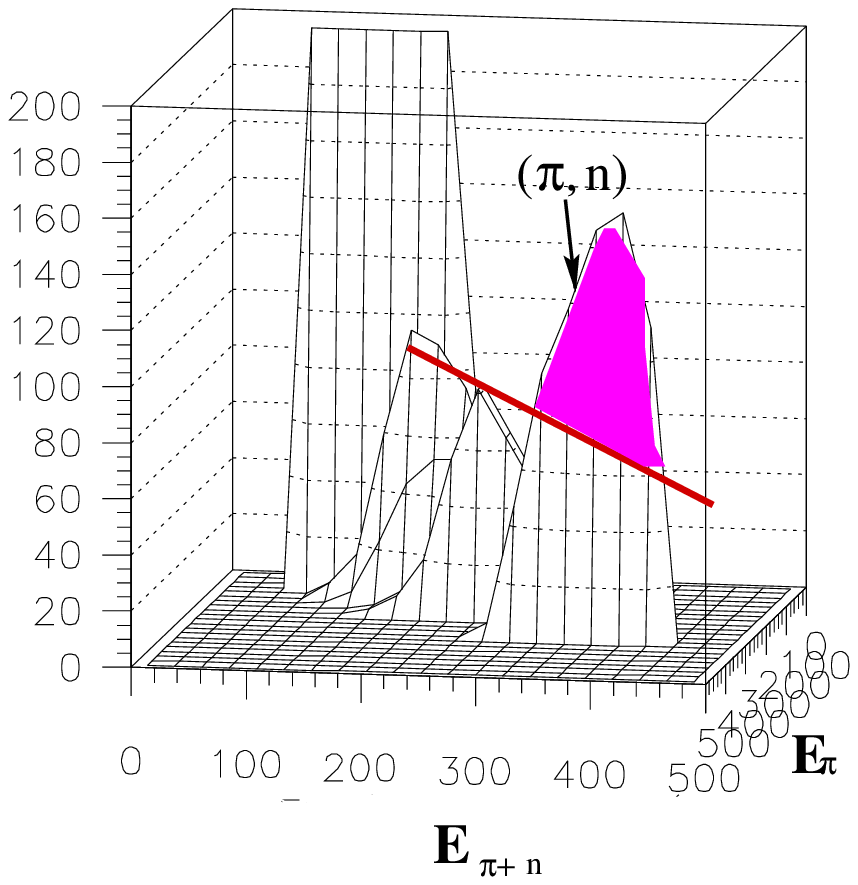} \hspace*{5mm}
\includegraphics[width=3.5cm,bb=157 322 443 613]{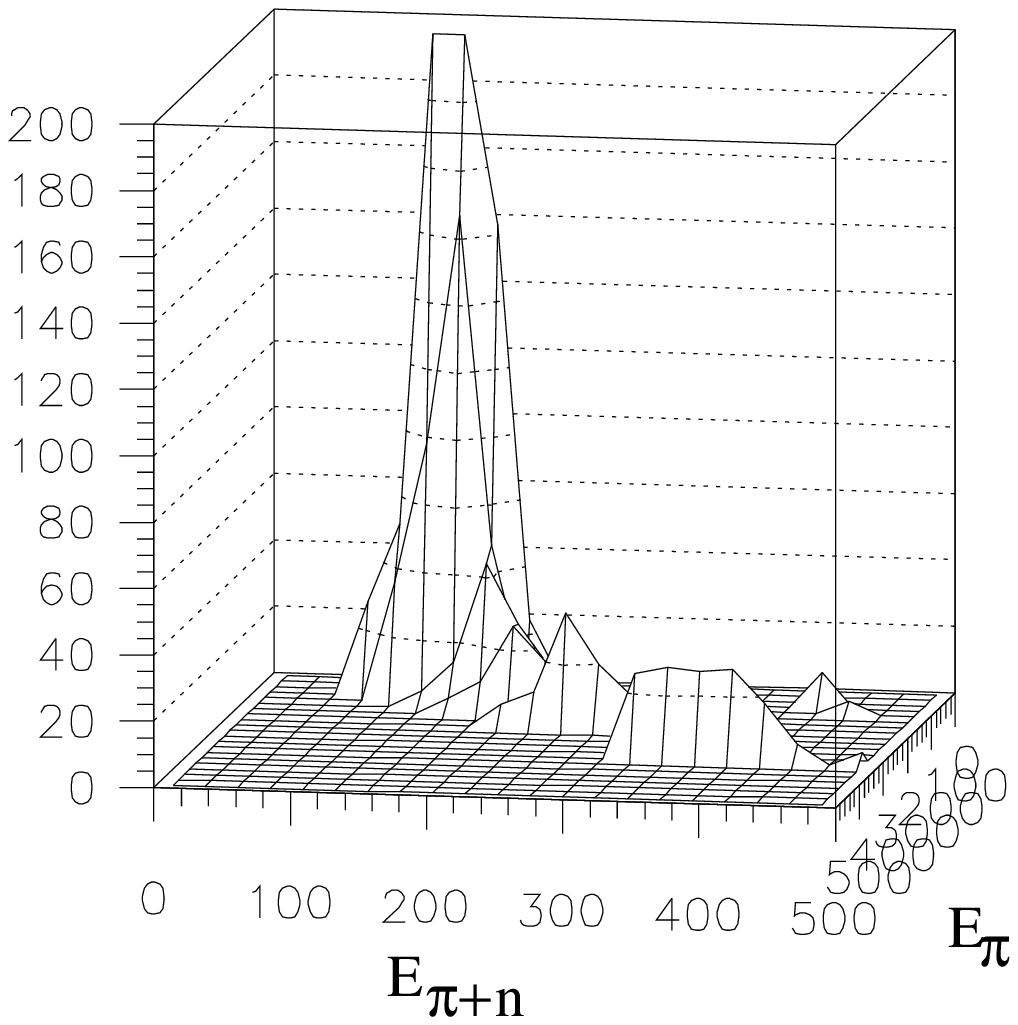} \hspace*{10mm}
\includegraphics[width=3.5cm,bb=120 416 334 650]{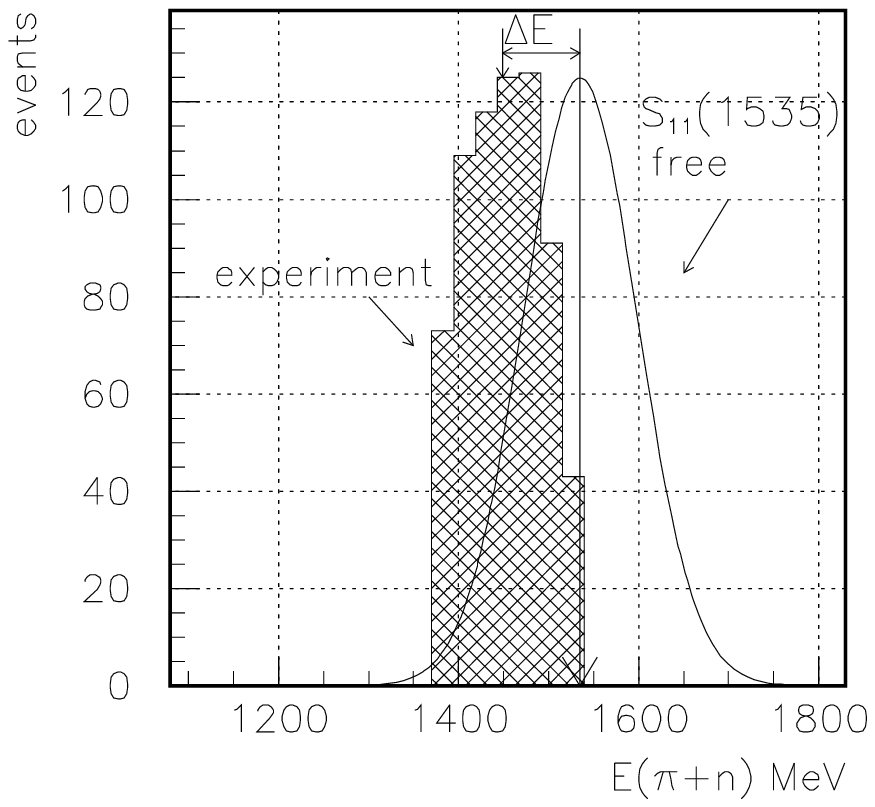}
}
\hspace*{4mm}$E_{\gamma \rm max} = 850$ MeV
\hspace*{8mm}$E_{\gamma \rm max} = 650$ MeV
\hspace*{14mm}$E_{\gamma \rm max} = 850$ MeV
\hfill \\
\parbox[t]{6cm}
{\caption
{Distribution  over the  total kinetic energy of the  $\pi^+n$ pairs
for the ``effect$+$background"  run (left panel) and for the
``background" run (right panel) obtained after unfolding raw spectra.}}
\hfill
\parbox[t]{5cm}
{\caption
{Distribution over the total energy of the  $\pi^+ n$ pairs after a
subtraction of the background. A distribution corresponding to a decay
of a free $S_{11}(1535)$ resonance is shown for a comparison.}}
\end{figure}
Therefore, an experimental $\beta$-resolution of the setup can be
directly inferred from the $\pi^0 \pi^0$ evens.  Then, using this
information and applying a statistical method of solving the inverse
problem described in [16], one can unfold the experimental spectrum,
obtain a smooth velocity distribution in the physical region $\beta_i
\le 1$ (Fig.~6) and eventually find a distribution of the particle's
kinetic energies $E_i = M_i [(1 - \beta^2_i)^{-1/2} - 1]$. Finding
$E_i$, we introduced corrections related with average energy losses of
particles in absorbers and in the detector matter.  Fig.~7 shows a
two-dimensional energy distribution of $\pi^+ n$ pairs obtained through
the above data evaluation for two runs at the beam energies of 650 and
850 MeV. At 650 MeV, Figure 7 shows a smooth decrease of the number
of $\pi^+ n$ pairs with the total energy of the pair.  At 850 MeV, a
resonance structure is observed in the kinematical region of the
$S_{11}(1535)$ resonance. We believe that such a structure is directly
related to $\eta$-nuclei formation and decay.

Of the most interest is the distribution of the $\pi^+ n$ events over
their total energy $E_{\rm tot} = E_n + E_\pi$. Subtracting a smooth
background (Fig.~7), we have found a 1-dimensional energy distribution
of the $\pi^+ n$ events presumably coming from bound $\eta$-mesons
decaying in the nucleus (Fig.~8). The experimental width of this
distribution is about 150 MeV. Its center lies by $\Delta E \approx 90
\pm 15$ MeV below the position of the $S_{11}(1535)$ resonance and even
below the threshold energy $m_\eta + m_N = 1485$ MeV, thus indicating a
presence of binding effects, i.e.\ a formation of $\eta$-mesic nuclei.

\section{Prospects}

Studies of $\eta$-mesic nuclei which lie at the
intersection of the nuclear physics and physics of hadrons promise us to
bring a new information important for both the fields.  $\eta$-mesic
nuclei provide a unique possibility to learn interactions of
$\eta$-mesons with nucleons and nucleon resonances in the nuclear
matter. Data on the self-energy of $S_{11}(1535)$ in the medium,
being interpreted in the framework of the chiral-symmetry models, can shed
more light on the problem of masses of free and bound hadrons [17].

%

\section*{Acknowledgements}

This work was supported the Russian Foundation for Basic Research,
grant N 99-02-18224.

\end{document}